\pdfoutput=1

\documentclass[
   ,final
  ]
  {aipproc}

\layoutstyle{8x11double}

\usepackage{amsmath,amssymb,amstext}
\usepackage{dsfont}
\usepackage{color}
\usepackage{slashed}

\newcommand{\be}{\begin{eqnarray}}
\newcommand{\ee}{\end{eqnarray}}

\begin{document}

\title{On the QCD phase diagram at finite chemical potential}


\author{Lisa M.~Haas}{
  address={Institut f\"ur Theoretische Physik, Universit\"at Heidelberg, Philosophenweg 16, 69120
    Heidelberg, Germany} ,altaddress={ExtreMe Matter Institute EMMI,
    GSI, Planckstr. 1, 64291 Darmstadt, Germany}
}

\author{Jens Braun}{
  address={Theoretisch-Physikalisches Institut, Universit\"at Jena, Max-Wien-Platz 1, 07743 Jena, Germany}
}

\author{Jan M.~Pawlowski}{
  address={Institut f\"ur Theoretische Physik, Universit\"at Heidelberg, Philosophenweg 16, 69120
    Heidelberg, Germany}
  ,altaddress={ExtreMe Matter Institute EMMI, GSI, Planckstr. 1, 64291 Darmstadt, Germany}
}

\begin{abstract}
 We present results for the phase diagram of QCD with two massless quark flavours 
 as obtained from a first-prin\-ci\-ples functional renormalisation group approach. In particular we
 compute order parameters for chiral symmetry breaking and quark
 confinement at vanishing and non-zero imaginary chemical
 potential. Our analytical and numerical results suggest a close
 relation between the chiral and the deconfinement phase transition. We
 discuss the properties of dual order parameters at imaginary and real
 chemical potential.
\end{abstract}

\maketitle


One of the unresolved problems in high energy physics is the structure of the QCD phase
diagram. This concerns in particular the transition(s) between
a deconfined and chirally symmetric phase with microscopic degrees of
freedom, quarks and gluons, to a phase of col\-our-neu\-tral macroscopic bound states, hadrons, with
broken chiral symmetry. The deconfinement transition is related to the breaking of the center symmetry of the gauge group and is driven by gluon dynamics. The
chiral transition is triggered by strong gluon-induced quark
interactions. We report on results of the QCD phase diagram and the
relation between the two phase transitions~\cite{Braun:2009gm}. The question whether both transitions are related is subject of an ongoing debate, see e.g. Refs.~\cite{Aoki:2009sc,Cheng:2009zi} for lattice 
and~\cite{Fukushima:2003fw,Ratti:2005jh,Schaefer:2007pw,Sakai:2009dv,Skokov:2010wb,Herbst:2010rf}
for model studies.

\paragraph{Approach}
We compute the QCD effective action with fRG techniques, which allow us to include all quantum fluctuations
step-wise at each momentum scale, for a recent overview see~\cite{Pawlowski:2010}. This is achieved by integrations
over small momentum shells, generating a flow from the microscopic
action in the UV towards the macroscopic action in the IR.
In our two-flavour calculation in the chiral limit~\cite{Braun:2009gm} we include the
Yang-Mills sector of QCD \cite{Braun:2007bx,Fischer:2008uz} and the matter sector \cite{Braun:2008pi,Gies:2002hq,Braun:2006jd,Schaefer:2004en} and couple them via
dynamic quark-gluon interactions. This approach has already been applied to
the chiral phase boundary in one-flavour QCD at finite chemical
potential ~\cite{Braun:2008pi}. The confining properties are
included via the full momentum dependence of the ghost and gluon
propagators \cite{Braun:2007bx,Fischer:2008uz,Braun:2010cy} and the matter sector incorporates dynamical mesonic
degrees of freedom.

\paragraph{Order parameters}
The order parameter for the deconfinement phase transition is the
Polyakov loop; it is proportional to the energy needed to put a
quark into the theory. In our study we implement the Polyakov loop
as defined in Ref.~\cite{Braun:2007bx,Marhauser:2008fz}. The order parameter for the
chiral transition is related to the quark condensate.

Recently so-called dual order parameters for the deconfinement phase
transition have been defined~\cite{Gattringer:2006ci}. This has been
extended in~\cite{Synatschke:2007bz} to any observable that transforms non-trivially under
center transformations and has been applied in
Refs.~\cite{Bilgici:2008qy,Bilgici:2009tx,Fischer:2009wc,Fischer:2010fx,Zhang:2010ui}. An
element of the center $Z(N_c)$ of the gauge group is given by $z=\mathds{1} e^{2 \pi i \theta_z}$, where
$\theta_z=0,1/3,2/3$ for $SU(3)$. It follows immediately that the sum over all center elements is zero in
the symmetric phase and non-zero in the broken phase. This means that
any observable that transforms non-trivially under center
transformations is an order parameter for confinement.

We have extended the above setting to imaginary chemical potential.
It can be incorporated in generalised boundary conditions of the quarks and
rewritten in terms of physical quarks with anti-periodic boundary conditions
\begin{eqnarray}\label{generalisedquarks}
\psi_{\theta}(x)=e^{2 \pi \text{i} \theta t/\beta} \psi(x)
\mbox{  with  } \psi(x)=\psi_{\theta=0}(x)
\end{eqnarray}
and $\beta=1/T$. Due to the periodicity in the angle $\theta$, we can Fourier decompose
general observables $\mathcal{O}_{\theta}=\langle \mathcal{O}[e^{2 \pi
\text{i} \theta t/\beta}\psi]\rangle$ which depend on the quark fields
\begin{equation}
\mathcal{O}_{\theta}=\sum_{l \in \mathds{Z}} e^{2 \pi \text{i} l \theta} O_l.
\end{equation}
Under center transformations the $O_l$ are multiplied with a
center element, $O_l \to z^l O_l$ and hence are order parameters for confinement
as they are proportional to the sum over all center elements and thus
vanish in the symmetric phase. One example is
$\mathcal{L}_{\theta}=e^{2 \pi \text{i} \theta} \langle L \rangle$,
where $L$ is the Polyakov loop variable. In particular for $l=1$ we find
\begin{equation}\label{dualorderparameters}
\tilde{\mathcal{O}}=\int_0^1 d\theta e^{-2 \pi \text{i} \theta}
\mathcal{O}_{\theta}=\int_0^1 d\theta O_1=O_1.
\end{equation}
Thus the dual Polyakov loop in QCD is $\langle L\rangle$.

In general, observables $\mathcal{O}_{\theta}$ can either be
evaluated in QCD with anti-periodic quarks, see
e.g. Refs.~\cite{Gattringer:2006ci,Synatschke:2007bz,Bilgici:2008qy,Bilgici:2009tx,Fischer:2009wc,Fischer:2010fx,Zhang:2010ui},
or in QCD at imaginary chemical potential, QCD$_{\theta}$, with
$\theta-$dependent boundary conditions~\cite{Braun:2009gm}. 

The Dirac action with quark fields defined
in~\eqref{generalisedquarks} reads
\begin{equation}
\int \bar\psi_\theta \left( i \slashed D+i m \right)\psi_\theta,
\end{equation}
where $\slashed D=\slashed \partial-ig \slashed A$. 
This can be rewritten such that we obtain an additional term which has the same form as an
imaginary chemical potential $\theta$:
\begin{equation}
\int \bar\psi\left( i \slashed D +i m -2 \pi \frac{1}{\beta} \gamma_0 \theta\right)\psi.
\end{equation}

Imaginary and real quark chemical potential are related via $\theta=-i \mu
\beta/2 \pi$. The effective action of QCD is then periodic under
a transformation of $\theta \to \theta+\theta_z$:
$\text{QCD}_{\theta}=\text{QCD}_{\theta+\theta_z}$, as the transformation of $\theta$ is canceled by the center
transformations of the fields. Observables $\mathcal{O}_{\theta}$ related to the
effective action show the same periodicity, namely the Roberge-Weiss (RW)
periodicity. However, this also means that if RW periodicity is not broken explicitly,
all $O_l$ vanish.
\begin{figure}
  \includegraphics[height=.3\textheight]{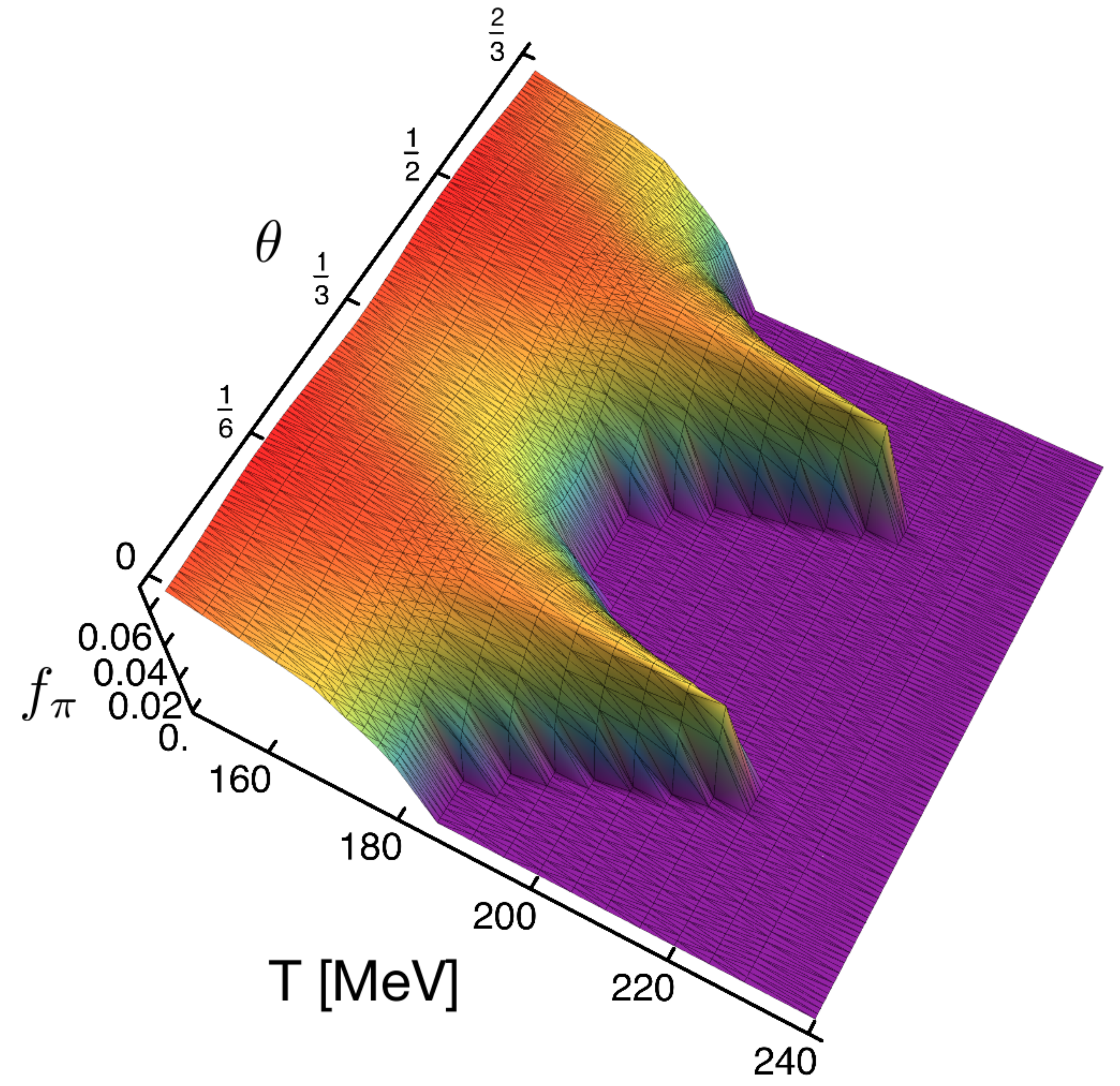}\label{fPi3d}
  \caption{The pion decay constant as a function of imaginary chemical
  potential and temperature.}
\end{figure}
The QCD phase diagram at imaginary chemical potential shows a
smooth transition until $\theta=1/6$ but then it displays a discontinuity: the Polyakov loop RW phase transition
 at $T_{\text{RW}}$ \cite{Roberge:1986mm}.

\paragraph{Presence of a fixed background field}
The RW periodicity of the generating functional is broken in the
presence of a current $J$. The dual observables $O_l$ then no longer vanish and as
derived above, serve as order parameters for confinement.
This is implemented by a $\theta$-independent gauge-field background $\varphi$. One example of the resulting order parameters is the dual density, which is proportional to the
logarithm of the generating functional. Therefore it grows like $T^3$ at high
temperatures as it is proportional to the first moment of
the grand canonical potential. Integration by parts
  yields the fermionic pressure difference  $\Delta P(T,\theta)= P(T,\theta)-P(T,0)$.

As the $\mathcal{O}_{\theta}$ are observables in different theories,
distinguished by the boundary condition, the $\tilde{\mathcal{O}}$
vanish only if $\text{QCD}_{\theta}$ is in the center symmetric phase
for all boundary conditions.

\begin{figure}
  \includegraphics[width=.41\textwidth]{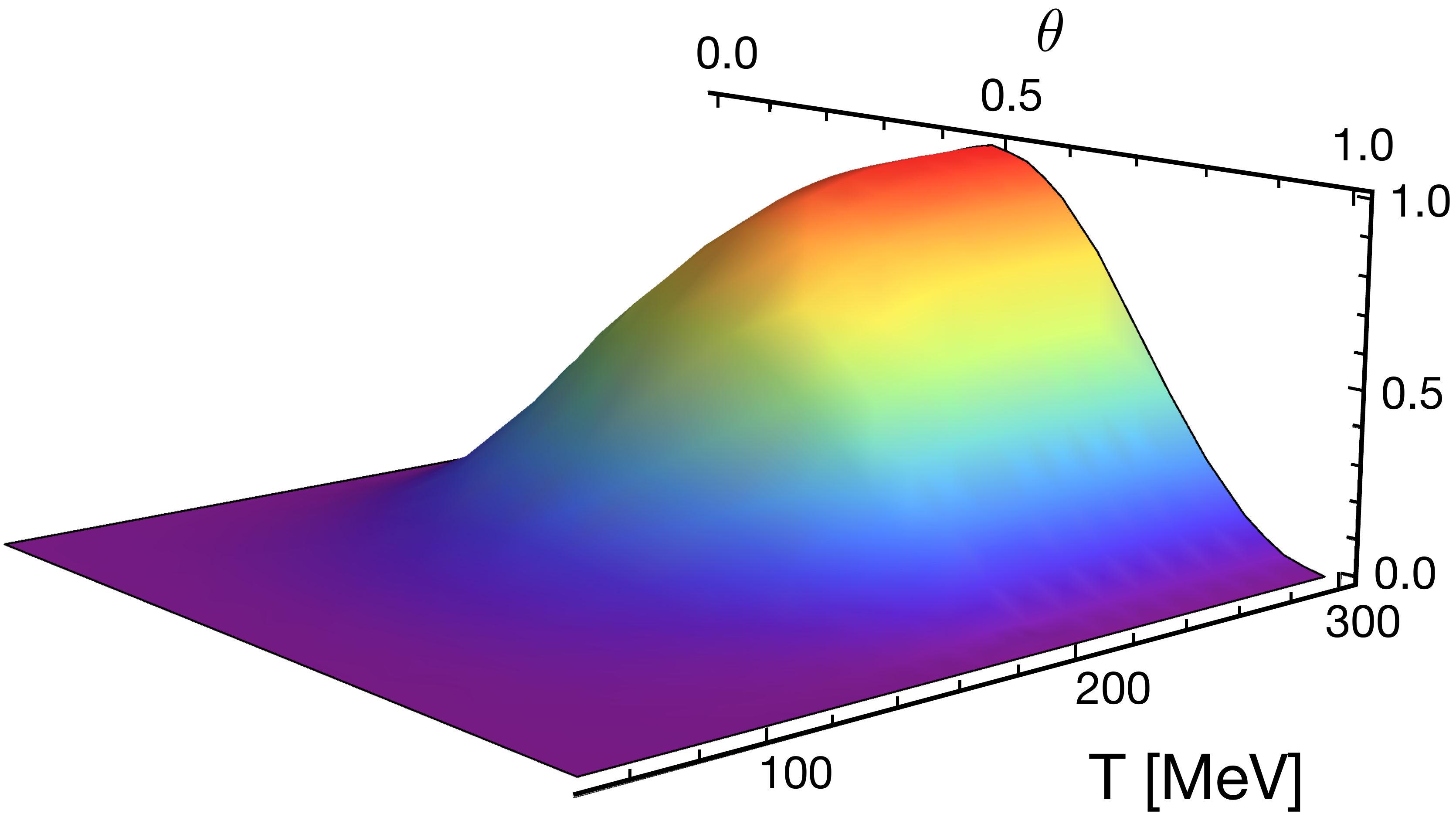}\label{ndual}
  \caption{$\Delta P(T,\theta)$ as a function of
    temperature and imaginary chemical potential.}
\end{figure}

\begin{figure}
  \includegraphics[width=.4\textwidth]{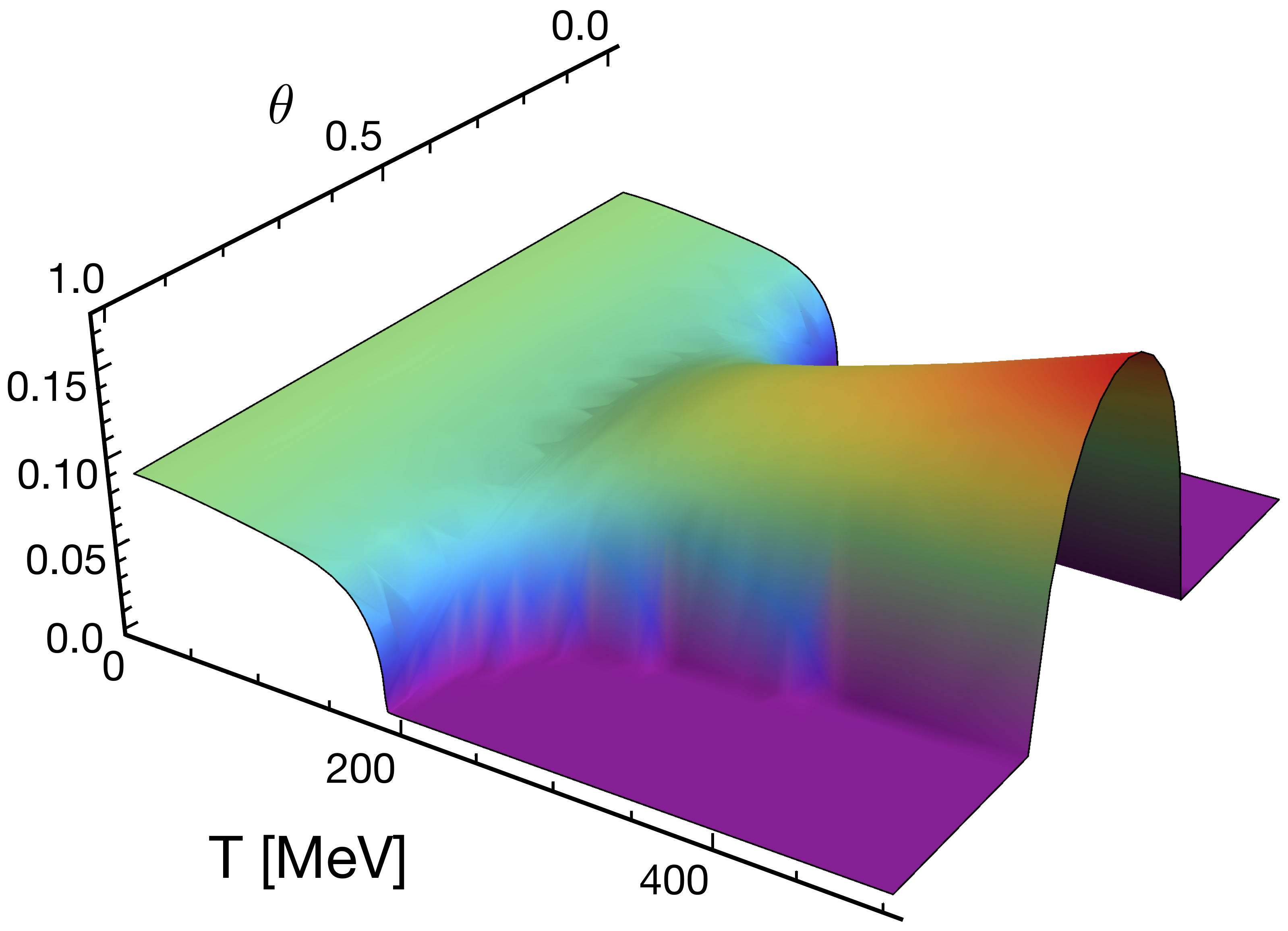}\label{mass}
  \caption{The dual quark mass parameter as a function of temperature and
    imaginary chemical potential.}
\end{figure}

Fig.~\ref{ndual} displays the
fermionic pressure difference in the presence of a $\theta$-independent gauge-field background $\varphi$. The RW symmetry is broken, however $\theta
\to \theta+1$ still holds. Instead of imaginary chemical potential,
one can think of this as an order parameter for different theories
which are distinguished by the value of the angle $\theta$.

\begin{figure}
  \includegraphics[height=.2\textheight]{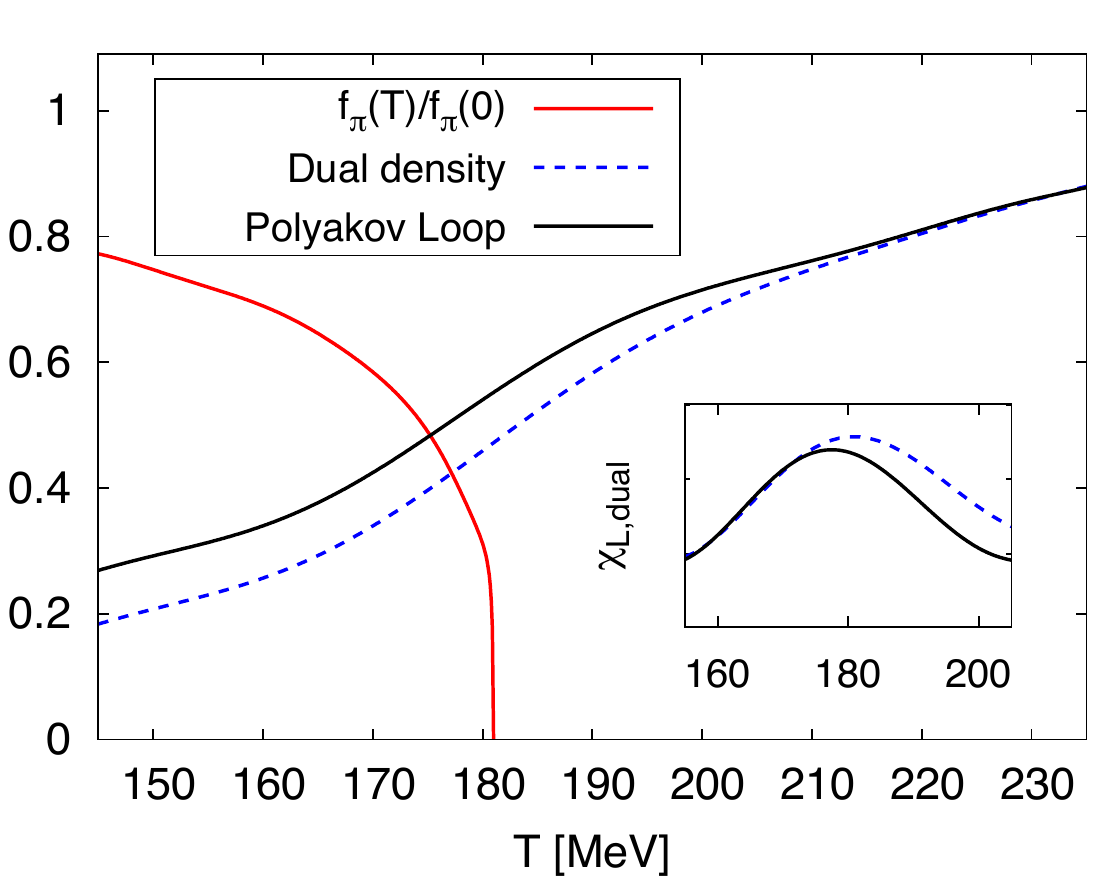}\label{order_parameters}
  \caption{The pion decay constant, the dual density and the Polyakov
    loop as functions of temperature, $\chi_L=\partial_T L$,
    $\chi_{\text{dual}}=\partial_T \tilde{n}$.}
\end{figure}

Another dual observable is the dual
quark mass parameter $\tilde{M}[\phi_J]$, see Eq.~\eqref{dualorderparameters} and $M_{\theta}[\phi_J]$ in Fig.~\ref{mass}. 
Due to the presence of a fixed background field it does not vanish in the broken
and in the symmetric phase. At $\theta=1/2$ closely above the
phase transition it increases with $\sqrt{T}$ and then linearly as the
quarks effectively have bosonic Matsubara frequencies at $\theta=1/2$. At $\theta=0$ the dual
quark mass is zero above the transition. For vanishing current it is related to the
pion decay constant in QCD: the slice of the 3D plot at $\theta=0$ is
the (normalised) line shown in Fig.~\ref{order_parameters}.

\paragraph{Results}
In our calculation we include the back-re\-ac\-tion of the matter sector
on the gauge sector. Moreover, we do not use input from, e. g., lattice
calculations to model the gauge dynamics. In other words, the gauge and the matter sector 
as well as their interplay are treated self-con\-sis\-tently within our approach.

At vanishing chemical potential we consider the order parameters of
the chiral and the deconfinement phase transition, the Polyakov loop,
the dual density and the pion decay constant, see Fig.~\ref{order_parameters}.

Above $T_c=180$ $\text{MeV}$ the pion decay constant vanishes and
chiral symmetry is restored. The dual density and the Polyakov loop both show a peak in their temperature
derivative at $\approx 178$ $\text{MeV}$. This provides a non-trivial consistency
check of our approximation as the Polyakov loop is computed from
gluonic correlation functions, whereas the dual density is computed
from matter correlation functions. We find that the chiral
and the deconfinement transition agree within a few $\text{MeV}$.

Fig.~\ref{fPi3d} displays the pion decay constant as a function of
imaginary chemical potential and temperature. For $T > T_{c,\chi}$
it vanishes and it is non-zero below $T_{c,\chi}$. The RW symmetry is
found. Moreover we find a second order phase transition as
expected in the chiral limit.

\begin{figure}
  \includegraphics[height=.19\textheight]{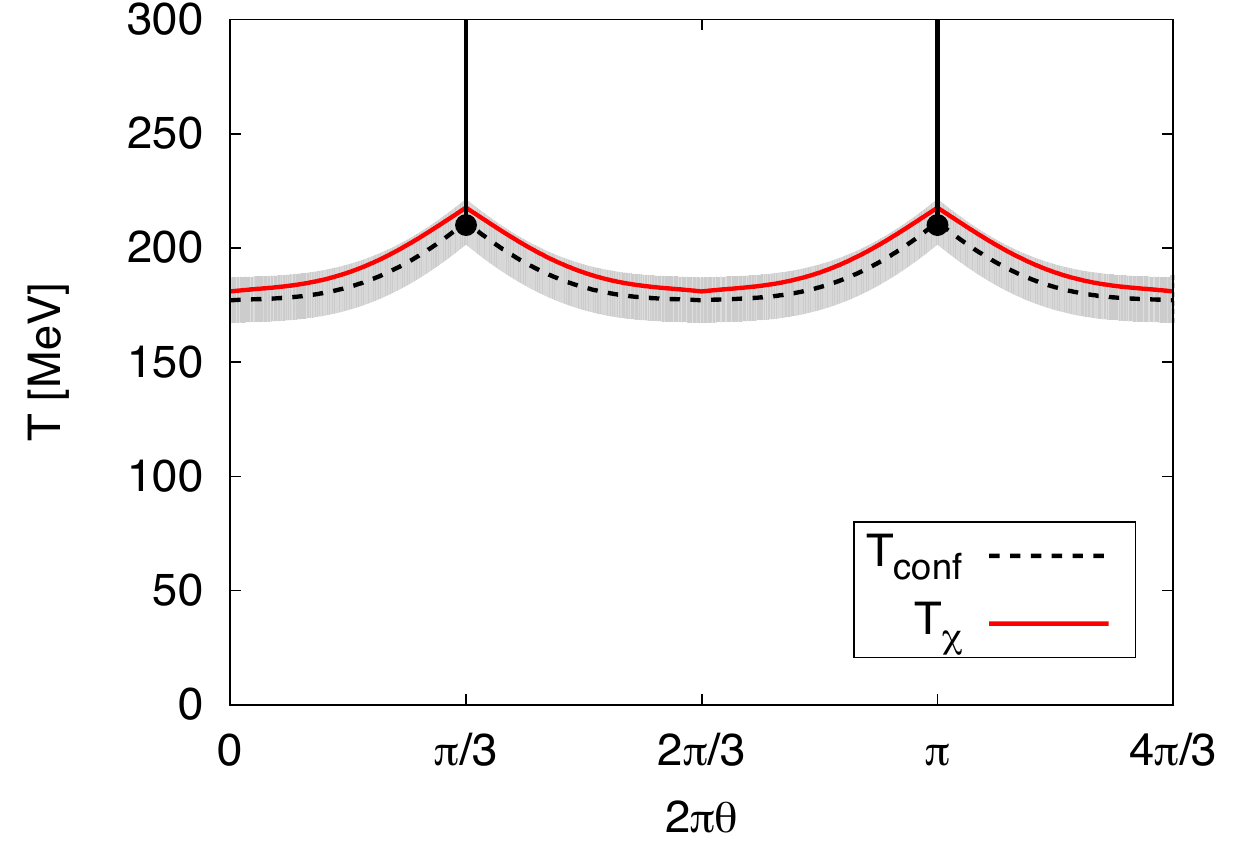}\label{phasediagram}
  \caption{The QCD phase diagram at imaginary chemical potential. The
    grey band represents the width of $\chi_L$. Black dots indicate the endpoints of the 
  Polyakov loop RW transitions.}
\end{figure}

Fig.~\ref{phasediagram} is a plot of the QCD phase diagram at
imaginary chemical potential. The chiral and the deconfinement transition agree within the width of the temperature
derivative of the Polyakov loop throughout the phase diagram. The
deconfinement transition occurs at lower critical temperatures than
the chiral transition. This was also found by lattice
computations~\cite{deForcrand:2002ci,D'Elia:2002gd}. First results
indicate that this persists at real chemical potential~\cite{Braun:2010}.


\paragraph{Acknowledgments}
This work is supported by Helmholtz Alliance HA216/EMMI.

\bibliographystyle{aipproc}

\bibliography{bibliography}

\end{document}